\documentclass{comjnl}

\usepackage{graphicx}
\usepackage{amsmath}
\usepackage{amssymb}
\usepackage{array}
\usepackage{color}
\usepackage{cite}
\usepackage{algorithm}
\usepackage{algorithmic}
\usepackage{multirow}
\usepackage{multicol}
\usepackage{booktabs}
\usepackage{url}
\usepackage{float}

\newcommand{\model}[0]{\texttt{DREIM}\xspace}

\usepackage{amsmath,amsfonts,bm}

\def\eqref#1{equation~\ref{#1}}

\def\1{\bm{1}}

\newcommand{\hb}{\mathbf{h}}

\newcommand{\wb}{\mathbf{w}}

\newcommand{\zb}{\mathbf{z}}

\newcommand{\Rb}{\mathbf{R}}

\newcommand{\Wb}{\mathbf{W}}
\newcommand{\Xb}{\mathbf{X}}

\newcommand{\cB}{\mathcal{B}}

\newcommand{\cD}{\mathcal{D}}
\newcommand{\cE}{\mathcal{E}}

\newcommand{\cG}{\mathcal{G}}

\newcommand{\cN}{\mathcal{N}}

\newcommand{\cS}{{\mathcal{S}}}

\newcommand{\cV}{\mathcal{V}}

\DeclareMathAlphabet{\mathsfit}{\encodingdefault}{\sfdefault}{m}{sl}
\SetMathAlphabet{\mathsfit}{bold}{\encodingdefault}{\sfdefault}{bx}{n}

\begin{document}

%
\title[Finding Influencers in Complex Networks]{Finding Influencers in Complex Networks: An Effective Deep Reinforcement Learning Approach}
\author{Changan~Liu}
\affiliation{Shanghai Key Laboratory of Intelligent Information
Processing, School of Computer Science, Fudan University, Shanghai 200433, China}
\authorstar{Changjun~Fan}
\affiliation{College of Systems Engineering, National University of Defense Technology, Changsha, China} 
\authorstar{Zhongzhi~Zhang}\email{fanchangjun@nudt.edu.cn, zhangzz@fudan.edu.cn}
\affiliation{Shanghai Key Laboratory of Intelligent Information
Processing, School of Computer Science, Fudan University, Shanghai 200433, China}

\shortauthors{C. Liu, C. Fan and Z. Zhang}


\begin{abstract}
Maximizing influences in complex networks is a practically important but computationally challenging task for social network analysis, due to its NP-hard nature. Most current approximation or heuristic methods either require tremendous human design efforts or achieve unsatisfying balances between effectiveness and efficiency. Recent machine learning attempts only focus on speed but lack performance enhancement. In this paper, different from previous attempts, we propose an effective deep reinforcement learning model that achieves superior performances over traditional best influence maximization algorithms. Specifically, we design an \textit{end-to-end} learning framework that combines graph neural network as the \textit{encoder} and reinforcement learning as the \textit{decoder}, named \model. Through extensive training on small synthetic graphs, \model outperforms the state-of-the-art baseline methods on very large synthetic and real-world networks on solution quality, and we also empirically show its linear scalability with regard to the network size, which demonstrates its superiority in solving this problem. 
\end{abstract}

\maketitle

\section{Introduction}

Social networks refer to a relatively stable system formed by various interactive relationships among individual members of society. Influence maximization problem is an important issue for social networks analysis, which has wide spread application in practice, including word-of-mouth marketing, crowd mobilization and public opinion monitoring~\cite{nguyen2016stop,chen2010scalable}. This problem can be formally described as finding out $k$ seeds~(influencers) to maximize their influences under certain propagation model, e.g., the independent cascade model (IC) or the linear threshold model (LT)~\cite{Kempe}.

Traditional attempts towards this problem can be categorized into two types~\cite{Tang2018sigmod,newchen2010icdm,newcheng2013cikm,newcheng2014SIGIR,newcohen2014cikm,newgalhotra2015asimWWW,newgalhotra2016sigmod,newgoyal2011icdm,newGoyal2011VLDB,newJung2021icdm,newLee2014www,newNaoto2014AAAI,newSong2015parallel,newTang2017ASONAM,newzhou2013icdm,newzhou2014www,leskovec2007costcelf,2012Maximizing,tang2014influence,tang2015influenceIMM,Kempe,bucur2016evelutionIM,centralityheuristic1,centralityheuristic2,2009Efficient}. First are approximate algorithms~\cite{leskovec2007costcelf,2012Maximizing,tang2014influence,tang2015influenceIMM}, which are bounded with theoretical guarantees, and can be applied to medium sized graphs. However the design of these algorithms often requires extensive expert error-and-trials and their scalability often suffers. Second ones are heuristic methods, including random heuristics ~\cite{Kempe}, evolution-based ones~\cite{bucur2016evelutionIM}, e.g., genetic algorithms, centrality-based ones~\cite{centralityheuristic1,centralityheuristic2} and degree discount heuristics~\cite{2009Efficient}. These methods are often fast and easy to design, and they can usually produce a feasible solution in a short time.They do not, however, offer any theoretical assurances, and the quality of the solutions can be very poor in some circumstances. Note that both approximate and heuristic methods are ad hoc in nature, with little cross-scenario flexibility.

To our best knowledge, the advantages of data-driven models such as deep reinforcement learning have not been well exploited for tackling the influence maximization problem. Recently, some works have proposed that reinforcement learning can be used to solve the combinatorial optimization problems on graphs~\cite{dai2017learning,manchanda2020gcomb,li2018guidedtreesearch}. The intuition behind these works is that the model can be trained in a graph distribution $\cD$, and the solution set  for a new graph can be obtained using the trained model. Dai et al.~\cite{dai2017learning} first implemented this idea. They applied this idea to solve the traditional combinatorial optimization problems such as minimum vertex coverage and maximum coverage problem. However, their method is difficult to apply to large-scale graphs. Later on, this method was improved by Li et al.~\cite{li2018guidedtreesearch}. Akash et al.~\cite{manchanda2020gcomb} first use this idea to solve the problem of influence maximization. Although the calculation speed has been improved, the performance of their algorithm i.e., selecting k seed nodes, and the proportion of nodes finally activated, is not better than the traditional best approximation algorithm IMM. Moreover, their model is trained via supervised learning, however, in real life, providing labels for supervised training is time-consuming and laborious. And supervised training will limit the ability of their model to obtain higher-quality solutions.

Inspired by current existing machine learning attempts in solving combinatorial optimization problems, here, we design \textbf{\model} (\textbf{D}eep \textbf{RE}inforcement learning for \textbf{I}nfluence \textbf{M}aximizations), an \textit{end-to-end} deep reinforcement learning based influence maximization model. More concretely, \model incorporates the graph neural network to represent nodes and graphs, and the Q-learning technique to update the trainable parameters. We train \model with large amounts of synthetic graphs, and the learned strategy could be applied on much larger instances, including both synthetic networks and real-world ones. \model achieves better solution quality than current state-of-the-art methods, for example, when selecting 100 seed nodes from the Facebook network, \model activates 26700 nodes while IMM activates 20180 nodes. Meanwhile, \model can also be very efficient through a \textit{batch nodes selection} strategy.

In summary, we make four main contributions:
\begin{itemize}
\item We formulate the seed nodes selection process of influence maximization (IM) problem as a Markov decision process.
\item We present an \textit{end-to-end} deep reinforcement learning framework to solve the classical IM problem. We design a novel  state representation for reinforcement learning by inducing a virtual node, which can capture system state information more accurately.
\item We propose a reasonable and effective termination condition of reinforcement learning, leading to the superior generalization capacity of \model.
\item We evaluate \model through extensive experiments on different sizes of large graphs. Our results demonstrate that \model is effective and efficient as well as its linear scalability on large networks.
\end{itemize}

The remaining parts are organized as follows. We systematically review related work in Section 2. After that, we present the preliminaries and problem formalization in Section 3. We then introduce the details of \model~architecture in Section 4. Section 5 presents the evaluation of \model on both large synthetic graphs and real-world networks. In section 6, we first intuitively discuss the policies learned by \model and then discuss the effect of our novel Q-learning setting. Finally, we conclude the paper in Section 7. 

\section{RELATED WORK}\label{related work.sec}

\subsection{Influence maximization}
Domingos and Richardson~\cite{richardson2002mining_imfirst1,domingos2001mining_imfirst2} perform the first study of influence maximization problem, and Kempe et al.~\cite{Kempe} first formulate the problem as a discrete optimization problem. Since then influence maximization has been extensively studied~\cite{Tang2018sigmod,newchen2010icdm,newcheng2013cikm,newcheng2014SIGIR,newcohen2014cikm,newgalhotra2015asimWWW,newgalhotra2016sigmod,newgoyal2011icdm,newGoyal2011VLDB,newJung2021icdm,newLee2014www,newNaoto2014AAAI,newSong2015parallel,newTang2017ASONAM,newzhou2013icdm,newzhou2014www,leskovec2007costcelf,2012Maximizing,tang2014influence,tang2015influenceIMM,Kempe,bucur2016evelutionIM,centralityheuristic1,centralityheuristic2,2009Efficient}. The existing solution methodologies can be classified into three categories.

\textbf{Approximation methods}. Kempe et al.~\cite{Kempe} proposed a hill-climbing greedy algorithm with a $1-1/e-\epsilon$ approximation rate, which uses tens of thousands of Monte Carlo simulations to obtain the solution set. Leskovec et al.~\cite{leskovec2007costcelf} then proposed a cost-effective delayed forward selection (CELF) algorithm. Experimental results show that the execution time of CELF is 700 times faster than that of the greedy algorithm. Borgs et al.~\cite{2012Maximizing} Borgs et al. proposed a IM sampling method called RIS. A threshold is set to determine how many reverse reachable sets are generated by the network and the node that covers the most of these sets is selected. TIM/TIM+~\cite{tang2014influence} greatly improves the efficiency of~\cite{2012Maximizing}, and is the first RIS-based approximation algorithm that achieves efficiency comparable to that of heuristic algorithms. Later, IMM~\cite{tang2015influenceIMM} employs the concept of martingale to reduce computation time while retaining TIM's $1-1/e-\epsilon$ approximation guarantee. According to a benchmark study~\cite{arora2017debunking}, IMM is known as the state-of-the-art approximation algorithm for solving IM problem.In addition, Tang et al.~\cite{Tang2018sigmod} proposed OPIM to improve interactivity and flexibility for a better online user experience. The disadvantages of most approximation methods are that they suffer from scalability problems and rely heavily on expert knowledge.

\textbf{Heuristic methods}. Unlike approximation methods, heuristic don't give any worst-case bound on the influence spread. These methods include the random heuristic~\cite{Kempe}, which randomly selects seed nodes; the centrality-based heuristic~\cite{centralityheuristic1,centralityheuristic2}, which selects nodes with high centrality; the evolution-based heuristic~\cite{bucur2016evelutionIM} and the degree-discounting heuristic proposed by Chen et al.~\cite{2009Efficient} which has similar effectiveness to the greedy algorithm and improves efficiency.

\textbf{Machine learning methods}. Akash et al.~\cite{manchanda2020gcomb} first leveraged reinforcement learning method to solve the influence maximization problem. Their model includes a supervised component that uses the greedy algorithm to generate solutions to supervise the neural network. However, generating the training data is a big challenge and supervised learning often suffers from the over-fitting issue.

\subsection{Graph representation learning}
Graph representation learning (GRL) tries to find $d$-dimensional $(d\ll{|\cV|})$ dense vectors to capture the graph information. The obtained vectors can be easily fed to downstream machine learning models to solve various tasks like node classification~\cite{sen2008collectivenodeclassification}, link prediction~\cite{liben2007linkprediction}, graph visualization~\cite{herman2000graphvisualization}, graph property prediction~\cite{fan2019drbc}, to name a few. $\mathrm{GNNs}$ are among the most popular graph representation learning methods which adopt a massage passing paradigm~\cite{kipf2017semisupervised,hamilton2017inductive}. Let $\hb_v^{(l)}$ denote the embedding vector of node $v$ at layer $l$, a typical $\mathrm{GNN}$ includes two steps: (\romannumeral1) Every node aggregates the embedding vectors from its neighbors at layer $l-1$; (\romannumeral2) Every node updates its embedding through combining its embedding in last layer and the aggregated neighbor embedding:
\begin{equation}
\label{eq:1}
\hb_{\cN(v)}^{(l)}=\operatorname{AGGREGATE}\left(\left\{\hb_{u}^{(l-1)}, \forall u \in \cN(v)\right\}\right),    
\end{equation}
\begin{equation}
\label{eq:gnn}
\hb_{v}^{(l)}=\beta\left(\Wb^{(l)} \cdot \operatorname{COMBINE}\left(\hb_{v}^{(l-1)}, \hb_{\cN(v)}^{(l)}\right)\right),    
\end{equation}
where $\cN(\cdot)$ denotes a set of neighboring nodes of a given node,  $W^{(l)}$ is a trainable weight matrix of the $lth$ layer shared by all nodes, and $\beta$ is an activation function, e.g., ReLU.

\subsection{Deep reinforcement learning} 
The reinforcement learning framework~\cite{sutton1988reinforcement} considers tasks in which the agent interacts with a dynamic environment through a sequence of observations, actions and rewards. Different from supervised learning, the agent in reinforcement learning is never directly told the best action under certain state, but learns by itself to realize whether its previous sequence of actions are right or not only when an episode ends. The goal of reinforcement learning is to select actions in a fashion that maximizes the long-term performance metric. More formally, it  uses a deep neural network to approximate the optimal state-action value function

\begin{equation}
\begin{split}
    &Q^{*}(s, a)=\\
    &\mathrm{max} _{\pi} \mathbb{E}\left[r_{t}+\gamma r_{t+1}+\gamma^{2} r_{t+2}+\ldots \mid s_{t}=s, a_{t}=a, \pi\right],    
\end{split}
\end{equation}
where $\gamma$ is the discounted factor, $\pi=P(a|s)$ is the behaviour policy, which means taking action $a$ at state $s$.

\subsection{Machine learning for combinatorial optimization}
In a recent survey, Bengio et al.~\cite{bengio2020machine4optimization} summarize three algorithmic structures for solving combinatorial optimization problems using machine learning: \textit{end to end} learning that treats the solution generation process as a whole~\cite{dai2017learning, bello2016neurale2e1,selsam2018learninge2e2}, learning to configure algorithms~\cite{bonami2018learningconfigure,wilder2019meldingconfigure2}, and learning in parallel with optimization algorithms~\cite{lodi2017learningalongside1,cappart2019improvingalongside2}. Dai et al.~\cite{dai2017learning} firstly highlight that it is possible to learn combinatorial algorithms on graphs using deep reinforcement learning. Then, Li et al.~\cite{li2018guidedtreesearch} and Akash et al.~\cite{manchanda2020gcomb} propose improvements from different aspects. Recently, Fan et al.~\cite{fan2020finder} have proposed an deep reinforcement learning framework to solve the key player finding problem on social networks. Most of these algorithms are based on two observations. First, although all kinds of social networks in real life are complex and changeable, the underlying generation models of these networks are unified, such as BA model~\cite{BA}, WS model~\cite{1998WS}, ER model~\cite{ER}, and powerlaw-cluster model~\cite{powerlaw-cluster}, etc. Second, the nodes in the solution set selected by the approximation algorithms should have similar characteristics, such as high betweenness centrality.

\section{PRELIMINARIES AND PROBLEM FORMALIZATION}
\label{graphs.sec}

In this section, we first introduce some preliminaries of the influence maximization problem and then give its formalization.

\subsection{Preliminaries}
\label{built.sec}

Let $\cG=(\cV,\cE,\Wb)$ be a social network, where $\cV$ is a set of nodes (users), $\cE$ is a set of edges (relationships), $|\cV|=N$ and $|\cE|=M$. $(u,v)\in{\cE}$ represents an edge from node $u$ to node $v$. Let $\Wb$ denote the weight matrix of edges indicating the degree of influence. $\cV_a$ denotes the active nodes set. 

\begin{itemize}
\item {\textbf{Seed node}}: A node $v\in{\cV}$ that is initially activated as the information source of the entire graph $\cG$. The set of seed nodes is denoted by $\cS$, and $\bar{\cS}$ is the complementary set of $\cS$.
\item {\textbf{Active node}}: A node $v\in{\cV}$ is regarded as active if it is a seed node ($v\in{\cS}$) or it is influenced by previously active node $u\in{\cV_a}$. 
\item {\textbf{Spread}}: The expected proportion of activated nodes after the process of influence propagation terminates, denoted as $\sigma(\cS)=\frac{|\cV_a|}{|\cV|}$.
\item {\textbf{Linear threshold model}}: LT model simulates the common herd mentality phenomenon in social networks. In this model, each node $v$ in a graph has a threshold $\theta_v$. Let $\cN(v)$ be the set of neighbors of node $v$ and $\cN^\alpha(v)$ be the set of activated neighbors of node $v$. For each node $u\in \cN(v)$, the edge $(u,v)$ has a non-negative weight $\wb(u,v)\leq 1$. Given a graph $\cG$ and a seed set $\cS$, and the threshold for each node, this model first activates the nodes in $\cS$. Then information starts spreading in discrete timestamps following the following rule. An inactive node $v$ will be activated if  $\sum_{u\in \cN^\alpha(v)}{\wb(u,v)}\ge\theta_{v}$. The newly activated nodes attempt to activate their neighbors. This process stops when no new nodes are activated.
\end{itemize}

\subsection{Problem formalization}
\label{topology.sec}

The influence maximization problem is formally defined as follows:
\begin{equation}
    \mathrm{argmax}_{|\cS|=k,\cS\subseteq \cV}\sigma(\cS),
\end{equation}
where $\cS$ is the solution set, $\sigma$ is the spread calculation function and $k$ is the budget.

We set the threshold of each node as a random real number between $0\sim1$. And for every node $v\in{\cV}$, we set the influence weight of its neighbors as $\frac{1}{|\cN_v|}$ where $|\cN_v|$ denotes the number of neighbors of node $v$. 
To our best knowledge, we are the first who utilize deep learning method to tackle the influence maximization problem under LT diffusion model.

\section{PROPOSED MODEL: DREIM}\label{model.sec}

In this section, we introduce the proposed model named \model. We begin with the overview and then introduce the architecture by parts as well as the training procedure. We also analyze the time complexity of \model in the end of this section.
\subsection{Overview}
\begin{figure*}[t!]
    \centering
    \includegraphics[width=0.95\textwidth]{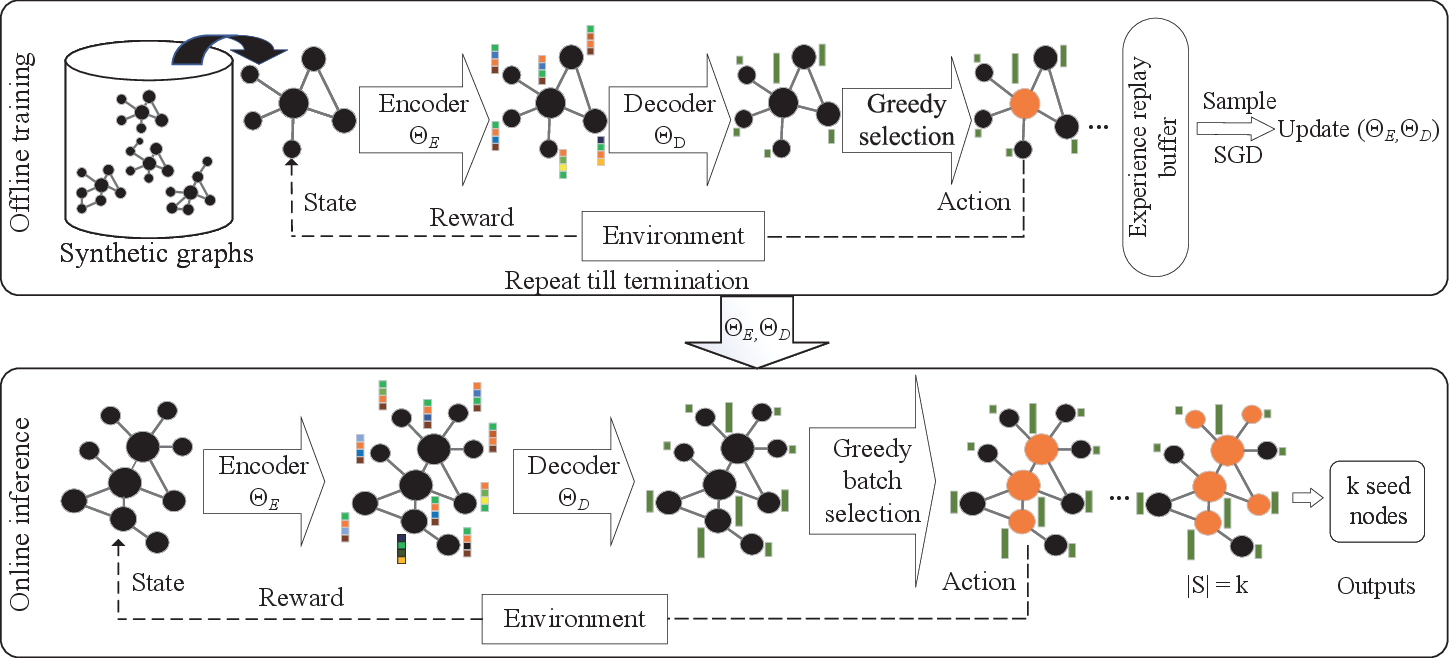}
    \caption{The pipeline of \model as a combination of the graph embedding and DQN process. The top half of the figure represents the training phase, and the bottom half is the testing phase. The colour bar denotes each node's embedding obtained after the encoding process and the green bar shows the Q-value of each node after the 2-layer MLP decoding process. Orange nodes indicate that they have been selected as seed nodes, and black nodes are not selected by \model.}
    \label{fig:pipeline}
\end{figure*}

The proposed deep learning model called \model combines graph neural network (GNN) and deep reinforcement learning (DRL) together in an \textit{end-to-end} manner. As illustrated in Fig. \ref{fig:pipeline}, \model has two phases: offline training and online inference. In the offline training phase (top), we train \model using synthetic graphs drawn from a network distribution $\cD$, like powerlaw-cluster model adopted here. At first we generate a batch of graphs scaling in a range, like $30\sim50$. Then we sample one (or a mini-batch) of them as the environment, and let \model interact with it. When the interaction process terminates, the experiences in the form of 4-tuple~$\left[S_{i}, A_{i}, R_{(i, i+n)}, S_{(i+n)}\right]$ will be stored in the experience replay buffer with a size of 500000. At the same time, the agent is getting more intelligent by performing mini-batch gradient descents over Eq.~$\left(\ref{eq:6}\right)$.
During the online inference phase (bottom), we applied the well-trained model to large synthetic and real-world networks with sizes scaling from thousands to millions. In order to decrease the computation time, we adopt a \textit{batch nodes selection} strategy which selects a batch of highest $Q$-value nodes at each adaptive step. We use the notion \model-X to denote different variants of \model, where X denotes the number of nodes selected at each step. \model refers to \model-1 and \model-All means we select k nodes as seed nodes at the very first step. 

There are three key parts in \model: (\romannumeral1) \textbf{Encoding}, which utilizes the GraphSAGE~\cite{hamilton2017inductive} architecture to learn the nodes' representations incorporating their structural information and feature information; (\romannumeral2) \textbf{Decoding}, which generates a scalar $Q$-value for 
each node; (\romannumeral3) \textbf{Greedy selection}, which adopts the $\epsilon$-greedy strategy in the training phase and the \textit{batch nodes selection} strategy in the inference phase, to make actions based on the $Q$ values of nodes.  In what follows, we will describe each part in detail.

\subsection{Encoder} Traditional hand-crafted features, such as node degree centrality, clustering coefficient, etc., can hardly describe the complex nonlinear graph structure. Therefore, we exploit the GraphSAGE~\cite{hamilton2017inductive} architecture to represent complex structure and node attributes using $d$-dimensional dense vectors. Alg.~\ref{alg:2} presents the pseudo code of GraphSAGE. The input node attributes vector $\Xb_v$ should include some raw structural information of a node. In this paper, we utilize a 2-tuple [out-degree, isseed] to set the input node attributes, where ``out-degree'' denotes the summation of outgoing edge weights of node $v$, and ``isseed'' denotes whether a node $v$ is already selected as seed node, i.e., this feature is set 1 if node $v$ is already a seed node and 0 if node $v$ is not a seed node. For the representation of the entire graph, we use the embedding of a virtual node that has connections to all the nodes in unique direction. 

\begin{algorithm}[t!]
\caption{Encoding algorithm}
\begin{algorithmic}[1]
\label{alg:2}
\REQUIRE $\cG=(\cV,\cE,\Wb)$; node attributes $\Xb_v\in{\Rb^{1\times c}}, \forall v \in \cV\cup\{s\}$;  iteration depth $L$; weight parameters $\Wb_{1} \in \Rb^{c \times d}, \Wb_{2} \in$
$\Rb^{d \times(d / 2)}, \Wb_{3} \in \Rb^{d \times(d / 2)}$
\ENSURE  Embedding vector $\zb_{v}, \forall v \in \cV \cup\{s\}$
\STATE Create a virtual node $s$ which connects all nodes in the graph, denoted as the graph state
\STATE Initialize $\hb_{v}^{(0)} \leftarrow \operatorname{ReLU}\left(\Xb_{v} \cdot \Wb_{1}\right), \hb_{v}^{(0)} \leftarrow \hb_{v}^{(0)} /\left\|\hb_{v}^{(0)}\right\|_{2}, \forall v \in \cV \cup\{s\}$

\FOR{$l = 1$ to $L$}
\FOR{$v \in {\cV}\cup\left\{s\right\}$}
\STATE{$\hb_{\cN(v)}^{(l-1)} \leftarrow \sum_{j \in \mathcal{N}(v)} \hb_{j}^{(l-1)}$}
\STATE{$\hb_{v}^{(l)} \leftarrow \operatorname{ReLU}\left(\left[\Wb_{2} \cdot \hb_{v}^{(l-1)}, \Wb_{3} \cdot \hb_{\mathcal{N}(v)}^{(l-1)}\right]\right)$}
\ENDFOR
\STATE{$\hb_{v}^{(l)} \leftarrow \hb_{v}^{(l)} /\left\|\hb_{v}^{(l)}\right\|_{2}, \forall v \in \cV \cup\{s\}$}
\ENDFOR
\STATE{$\zb_{v} \leftarrow \hb_{v}^{(K)}, \forall v \in \cV \cup\{s\}$}
\end{algorithmic}
\end{algorithm}

\subsection{Decoder} In the previous step, we obtain the embeddings, in which the virtual node's embedding can be regarded as the state while other nodes' embeddings as the potential actions. In the decoding step, we aim to learn a function that can transfer the state-action pair $\left(s, a\right)$ to a scalar value $Q\left(s, a\right)$. The scalar value indicates the expected maximal rewards after taking action $a$ given state s. 
In \model, the embeddings of state and actions are fed into a 2-layer MLP. We employ rectified linear unit  $\left(\mathrm{ReLU}\right)$ as the activation function. Formally, the decoding process can be defined as follows:
\begin{equation}
Q(s, a)=\Wb_{5}^{\top} \operatorname{ReLU}\left(\zb_{a}^{\top} \cdot \zb_{s} \cdot \Wb_{4}\right),    
\end{equation}
where $\Wb_{4} \in \mathbf{R}^{d \times 1}, \Wb_{5} \in \mathbf{R}^{d \times 1}$ are weight parameters between two neural network layers, $\zb_s$ and $\zb_a \in \mathbf{R}^{1 \times d}$ are the output embeddings for state and action respectively.
We present how we define the elements of $Q$-learning as follows:
\begin{itemize}
\item {\textbf{State}}: 
We create a virtual node to represent the state $s$. $s$ updates its embedding in the same way as other nodes do, i.e., message aggregation and combination.  
\item {\textbf{Action}}: Action is the process of adding a node $v \notin {\cS}$ to the solution set $\cS$.
\item {\textbf{Transition}}: When a node $v$ is selected to join the solution set $\cS$, the ``isseed'' attribute of it will change from 0 to 1.

\begin{figure}[t!]
  \centering
  \includegraphics[width=\linewidth]{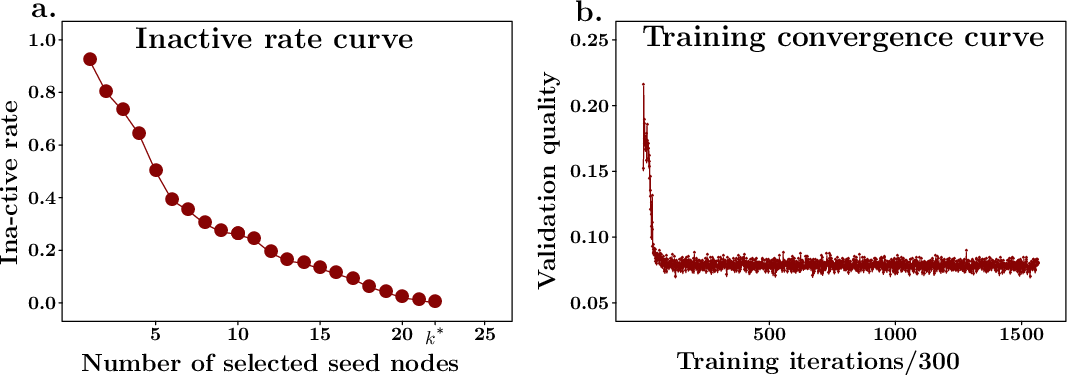}
  \caption{Training analysis of \model. (a). Remaining inactive rate decreases as the number of seed nodes increases. $k^*$ seed nodes are selected to activate all the nodes in a network. (b). The training process convergences fast measured by validation quality, i.e., area under the inactive rate curve.}
  \label{fig:convergence curve}
\end{figure}
\item {\textbf{Reward}}: After the agent makes an action, it will receive a feedback from the environment, which represents reward or punishment. In \model, we use negative inactive rate $-\frac{|\cV|-|\cV_a|}{|\cV|}$, i.e., $\sigma - 1$ to denote the reward after adding a node into $\cS$.  
\item {\textbf{Policy}}: Policy is the rule the agent obeys to pick the next action. In this work, we adopts different policies during training and inference, and we introduce it concretely in section~\ref{Greedy} 
\item {\textbf{Termination}}: The interaction process terminates when all nodes are activated. We choose this termination condition to improve \model's generalization ability. Previous works set the termination condition as $|\cS| = k$~\cite{manchanda2020gcomb,nipa}, we argue this will hinder the model's generalization ability. For example, when the budget $k = 10$, there will be 90 nodes left for a graph with 100 nodes, but there will be 990 nodes left for a graph with 1000 nodes, which will confuse the agent, since the terminal state is totally different for graphs with different sizes. 
\end{itemize}

\subsection{Greedy Selection} \label{Greedy}
Based on the above two steps, we adopt a greedy selection process to make actions based on the $Q$ values of nodes. In the training phase, since the model has not been trained well, we adopt the $\epsilon$-greedy strategy 
to balance exploration and exploitation. 
In the inference phase, since we have already trained \model well, we only exploit the learned $Q$ function to take the highest $Q$ value action. In order to speed up the solution set generation process, we also adopt a  \textit{batch nodes selection} strategy, which takes the top-$k$ nodes with the highest $Q$-value at each adaptive step. Extensive analysis shows that this strategy can bring a significant speed increase without compromising much on solution quality.
\begin{algorithm}
\caption{Training algorithm}
\begin{algorithmic}[1]
\label{alg:3}
\REQUIRE number of episodes $e$, replay buffer size $b$, 
\STATE Initialize \textbf{DQN} $\Theta$ and target \textbf{DQN} $\hat{\Theta}$
\STATE Initialize $n$-step experience replay buffer $B$
\FOR{episode =1 to $e$}
\STATE Draw random graphs from distribution $D$, like \textbf{powerlaw-cluster} model
\STATE Initialize the solution set $S$ as an empty set $\emptyset$
\FOR{t=1 to Termination}
\STATE Compute embeddings and $Q$ values
\STATE c $\leftarrow$ random number between 0 and 1
\STATE $v_{t}=\left\{\begin{array}{ll}\text {random node } v \in \bar{\cS}, &c < \varepsilon \\ \operatorname{argmax}_{v \in \bar{\cS}} Q(v, \cS;\Theta),& c \ge \varepsilon\end{array}\right.$
\STATE $\cS_{t+1}=\cS_{t} \cup v_{t}$, get reward $r_t$
\IF{$t \geq n$}
\STATE Add tuple $\left(s_{t-n}, a_{t-n}, r_{t-n, t}, s_{t}\right)$ to $\cB$
\STATE Sample random batch experiences from $B$
\STATE Perform stochastic gradient descents for $\Theta$
\STATE Every $C$ steps reset $\hat{\Theta}=\Theta$
\ENDIF
\ENDFOR
\ENDFOR 
\end{algorithmic}
\end{algorithm}

\subsection{Training and inference}
During training, there are two sets of parameters to be learned, including $\Theta_{\mathrm{E}}=\left\{\Wb_1, \Wb_2, \Wb_3\right\}$ and $\Theta_{\mathrm{D}}=\left\{\Wb_4, \Wb_5\right\}$. Thus total parameters to be trained can be denoted as $\Theta=\left\{\Theta_{\mathrm{E}},\Theta_{\mathrm{D}}\right\}$.
To avoid the inefficiency of Monte Carlo method~\cite{sutton1988reinforcement}, we adopt the Temporal Difference method~\cite{fittedQ} which leads to faster convergence. To better estimate the future reward, we utilize the $n$-step $Q$-learning~\cite{sutton1988reinforcement}.  
We use an experience replay buffer to increase the training data diversity. We not only consider the $Q$-learning loss like other works do~\cite{manchanda2020gcomb,nipa} but also the graph reconstruction loss to enhance topology information preservation. Alg. \ref{alg:3} presents the whole set of training process of \model. The loss function can be formalized as follows:
\begin{equation}
\begin{split}
    \operatorname{Loss}\left(\Theta\right)=&y-Q\left(s_{t}, a_{t} ; \Theta_{Q}\right)^{2}+\\
    &\alpha \sum_{i=1}^{N}\sum_{j=1}^{N} s_{i, j}\left\|z_{i}-z_{j} ; \Theta_{\text {E }}\right\|_{2}^{2},
\end{split}
\label{eq:6}
\end{equation}
where $y=r_{t, t+n}+\gamma \max _{a^{\prime}} \hat{Q}\left(s_{t+n}, a^{\prime} ; \hat{\Theta}_{Q}\right)$, $\Theta_{Q}=\left\{\Theta_{\mathrm{E}}, \Theta_{\mathrm{D}}\right\}$ and $\hat{\Theta}_{Q}$ is the parameters of the target network, which are only updated as $\Theta_{Q}$ every C iterations. $\alpha$ is a hyper-parameter to balance the two losses. $\gamma$ is the discounting factor. We validate \model using the metric \textit{Return}:
\begin{equation}
Return\left(v_{1}, v_{2}, \ldots, v_{k^*}\right)=\frac{1}{N} \sum_{k=1}^{k^*} \left(1 - \sigma\left(\left\{v_{1}, v_{2}, \ldots, v_{k}\right\}\right)\right),
\label{eq:7}
\end{equation}
where $k^*$ is the number of seed nodes needed to activate all the nodes in a network. \textit{Return} can be regarded approximately as the area under the inactive rate curve as shown in Fig. \ref{fig:convergence curve} (a).

\begin{table*}
\caption{Active rate comparison of different methods on synthetic networks for $k = 50$, $k = 100$, $k = 200$  (\%). All the results are the average value for 100 networks of the same scale.}\label{tab:1}
\resizebox{\linewidth}{!}{%
\begin{tabular}{@{}|c|c|c|c|c|c|c|c|c|c|@{}}
\toprule
\multirow{2}{*}{Scale} & \multicolumn{3}{c|}{$k = 50$}& \multicolumn{3}{c|}{$k = 100$}     & \multicolumn{3}{c|}{$k = 200$} \\ 
\cmidrule(l){2-10} 
&IMM&GCOMB&\model&IMM&GCOMB&\model&IMM&GCOMB&\model\\
\midrule
10000        &$39.84\pm0.39$ &$39.47\pm0.61$ &$\pmb{40.89\pm0.42}$ & $30.37\pm0.41$ & $29.43\pm0.25$ &$\pmb{30.41\pm0.32}$ & $60.14\pm0.44$ &$58.12\pm0.12$ &$\pmb{60.29\pm0.21}$\\ 
\midrule
            
20000       &$30.37\pm0.41$ & $29.43\pm0.25$ &$\pmb{30.41\pm0.32}$ & $39.28\pm0.41$ &$38.69\pm0.17$ & $\pmb{40.59\pm0.48}$ & $49.36\pm0.37$ &$48.36\pm0.41$ &$\pmb{49.98\pm0.41}$ \\ 
\midrule
            
50000       &$21.96\pm0.35$  &$20.61\pm0.34$ & $\pmb{21.98\pm0.21}$ & $29.24\pm0.32$ &$28.55\pm0.39$ & $\pmb{30.43\pm0.11}$ & $37.55\pm0.28$ &$37.33\pm0.46$ & $\pmb{37.86\pm0.16}$ \\ 
\midrule
            
100000       &$16.50\pm0.25$ &$16.39\pm0.38$ & $\pmb{17.56\pm0.37}$
            &$22.22\pm0.28$ &$21.99\pm0.16$ & $\pmb{23.72\pm0.15}$
            &$29.14\pm0.24$ &$28.81\pm0.16$ & $\pmb{29.89\pm0.40}$ \\ 
\midrule
            
500000      &$8.74\pm0.11$ &$8.75\pm0.46$ & $\pmb{9.22\pm0.68}$
            &$12.03\pm0.14$ & $12.16\pm0.24$ &$\pmb{12.53\pm0.18}$ & $16.08\pm0.12$ &$16.48\pm0.19$ & $\pmb{17.01\pm0.34}$\\ 
\bottomrule
\end{tabular}}
\end{table*}

In the inference phase, we adopt the \textit{batch nodes selection} strategy, i.e., instead of one-by-one iteratively selecting and recomputing the embeddings and $Q$ values, we pick a batch of highest-$Q$ nodes at each adaptive step.

\subsection{Complexity analysis}
The complexity of \model consists of two processes, i.e., training process and inference process.

\noindent\textbf{Training complexity.} The complexity of the training process depends on the training iterations, which is hard to be theoretically analyzed. Experimental results show that \model convergences fast, which indicates that we don't need many iterations for training. For example, as shown in Fig. \ref{fig:convergence curve} (b), we train \model on synthetic graphs with a scale range of $30\sim50$, \model convergences at the iteration about 60000 meaning that the training procedure is not much time-consuming.\\
\textbf{Inference complexity.} The inference complexity is determined by three parts, encoding, decoding and node selection, with a complexity of $O(M)$, $O(N)$, and $O(N\mathrm{log}N)$ respectively, which results in a total complexity of $O(M+ N+N\mathrm{log}N)$. Since most real-world social  networks are sparse, thus we can say \model has linear scalability concerning the network size.
Note that \model is once-training multi-testing, i.e., the training phase is just performed only once, and can be used for any input network in the inference phase.

\begin{figure}
    \centering
    \includegraphics[width = \linewidth]{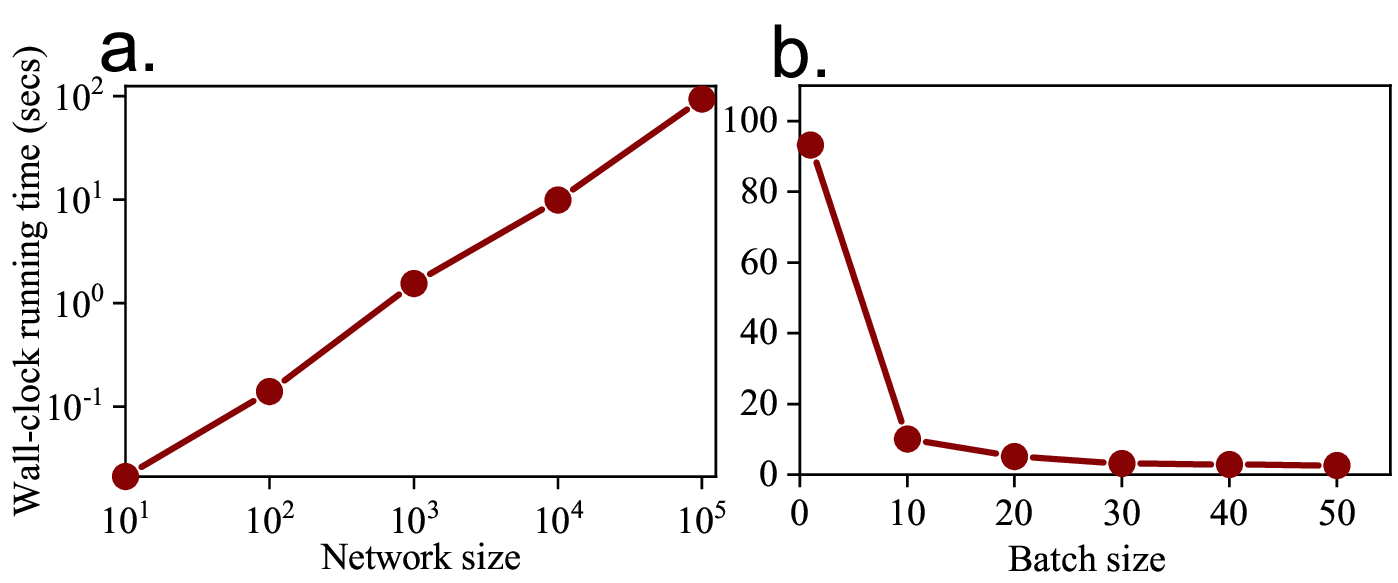}
    \caption{Wall-clock running time analysis. (a). Liner scalability of \model. 
    (b). For Facebook network with budget $k=50$, the wall-clock running time decreases quickly as the batch size increases.}
    \label{fig:scale}
\end{figure}
\begin{table}
  \caption{Basic statistics of real-world networks. $\gamma$ denotes the exponent of the power-law degree distribution. }
  \label{tab:statistics}
  \begin{tabular}{c|ccccc}
    \toprule[1.5pt]
    Network&\# Nodes&\# Edges&$\gamma$&Avg.Degree\\
    \midrule[1pt]
    Digg & 29.6K & 84.8K & 2.79&5.72\\
    Facebook & 63.3K & 816.8K & 2.43&25.77\\
    Youtube & 1.13M  & 2.99M & 2.14&5.27\\
    LiveJournal & 4.85M & 69M & 2.43&6.5\\
  \bottomrule[1.5pt]
\end{tabular}
\end{table}
\begin{figure}
  \centering
  \includegraphics[width=\linewidth]{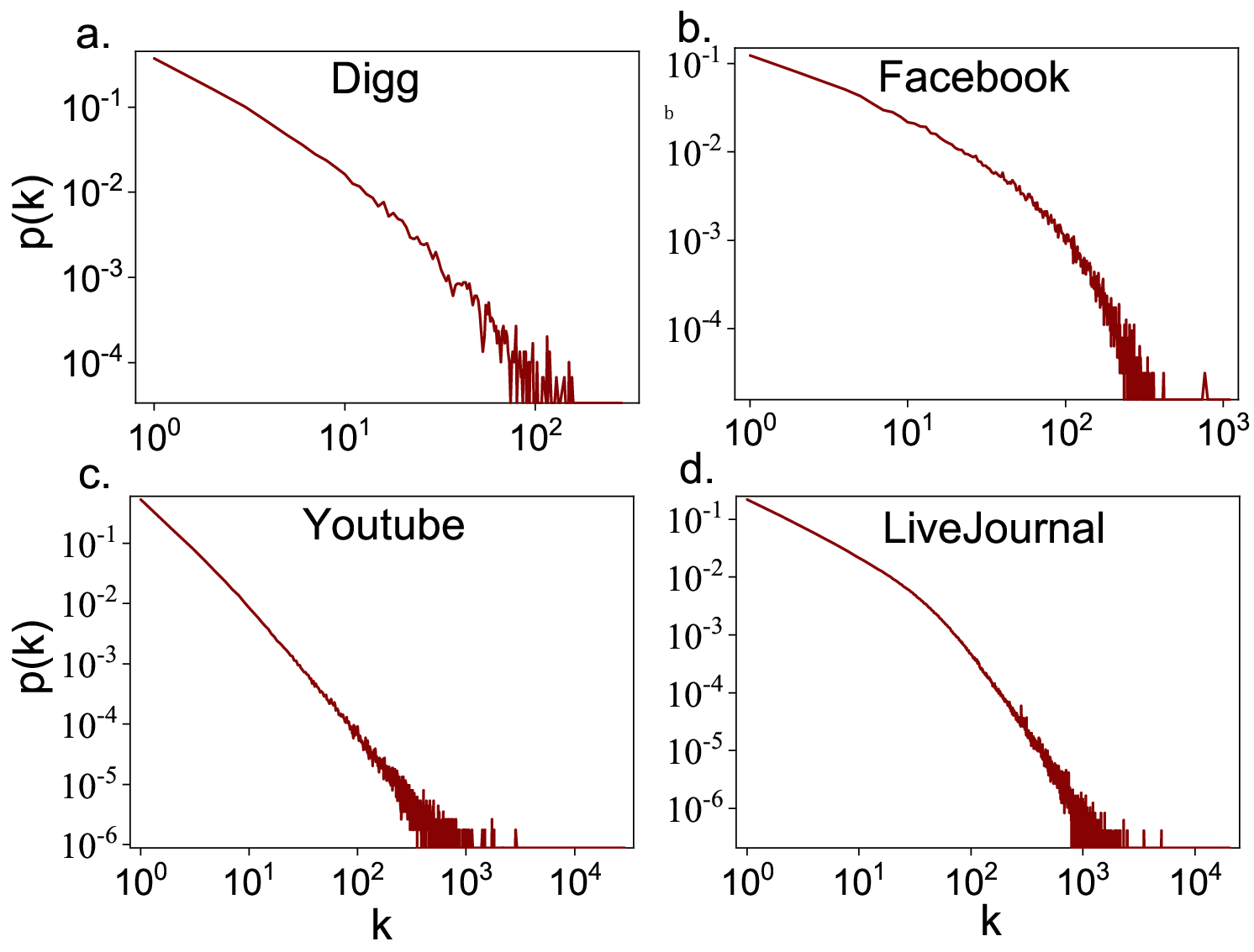}
  \caption{Degree distribution of real-world networks.}
  \label{fig:Degree Distribution}
\end{figure}
\section{EXPERIMENT}
We train and validate \model using synthetic graphs generated by the powerlaw-cluster model as it captures two important features of the vast majority of real-world networks, i.e., small-world phenomenon~\cite{1998WS} and power-law degree distribution~\cite{BA}. And we test \model on both large synthetic and real-world social networks of different scales. To our best knowledge, this is the first work that considers learning for IM problem under the LT diffusion model, moreover, \model exceeds the spread quality of all the baselines.
\subsection{Experimental setup}
\subsubsection{Baselines}
We compare the effectiveness and efficiency of \model with other baselines. IMM is the state-of-the-art approximation method according to a benchmarking study~\cite{arora2017debunking}. Previous work GCOMB~\cite{manchanda2020gcomb} tries to solve the IM problem using reinforcement learning method and obtains similar performance to IMM. So we compare \model with IMM and GCOMB. We use the code of IMM and GCOMB shared by the authors. For IMM, we set $\epsilon = 0.5$ throughout the experiment as suggested by the authors. For GCOMB, to make it comparable, we change the diffusion model from IC to LT. We train and validate GCOMB on subgraphs sampled from Youtube by randomly selecting 30\% of its edges, and we test GCOMB on the remaining subgraph of Youtube and the entire graphs of other networks. To reduce the randomness of the LT diffusion model, we ran 10000 diffusion processes and report the average value as the final spread.
\subsubsection{Datasets}
We adopt the powerlaw-cluster model to generate synthetic networks with different scales. 
Table \ref{tab:statistics} summarize the basic statistics of real-world networks~\cite{snapnets} and Fig. \ref{fig:Degree Distribution} illustrates the degree distribution of them.

\subsubsection{Evaluation metrics}
To evaluate \model quantitatively, we consider the following three aspects to report \model's effectiveness and efficiency.

\noindent \textbf{Spread quality.} We adopt the metric \emph{active rate}, i.e., the proportion of active nodes to total nodes under a specific seed nodes budget $k$ to compare the spread quality with baselines.\\
\noindent \textbf{Scalability.} As analyzed before, the inference complexity of \model is $O(kM)$, where $k$ is the given budget and $M$ is the number of edges. We test \model on large networks scaling up to millions of nodes. We can see in Fig. \ref{fig:Active Rate} that \model can effectively solve the problem of networks size up to 4.85 million.\\ 
\textbf{Time complexity.} We theoretically compare the time complexity of \model and all the baselines.

\begin{table*}
\caption{Active rate comparison of different \model variants, i.e., using distinct batch sizes on synthetic networks for $k = 50$, $k = 100$, $k = 200$  (\%). All the results are the average value for 100 networks of the same scale.}\label{tab:4}
\resizebox{\linewidth}{!}{
\begin{tabular}{@{}|c|c|c|c|c|c|c|c|c|c|@{}}
\toprule
\multirow{2}{*}{Scale} & \multicolumn{3}{c|}{$k = 50$}& \multicolumn{3}{c|}{$k = 100$}     & \multicolumn{3}{c|}{$k = 200$} \\ 
\cmidrule(l){2-10} 
&\model-1&\model-10&\model-All&\model-1&\model-10&\model-All&\model-1&\model-10&\model-All\\ 
\midrule
10000         &$40.89\pm0.42$ & $\pmb{41.70\pm0.62}$&$38.72\pm0.67$
             &$30.41\pm0.32$ & $\pmb{34.19\pm0.21}$&$29.62\pm0.18$
             &$\pmb{60.29\pm0.21}$ & $59.16\pm0.41$&$60.20\pm0.62$ \\ 
\midrule
            
20000       &$30.41\pm0.32$ & $\pmb{34.19\pm0.21}$&$29.62\pm0.18$
            & $\pmb{40.59\pm0.48}$ & $38.73\pm0.73$&$39.25\pm0.42$
            &$\pmb{49.98\pm0.41}$ & $49.76\pm0.29$&$47.71\pm0.32$ \\ 
\midrule
            
50000       & $\pmb{21.98\pm0.21}$ & $20.25\pm0.56$&$21.28\pm0.53$
            & $\pmb{30.43\pm0.11}$ & $29.01\pm0.65$&$28.99\pm0.28$
            & $37.86\pm0.16$ & $\pmb{37.94\pm0.36}$&$36.63\pm0.26$ \\ 
\midrule
            
100000       & $\pmb{17.56\pm0.37}$ & $16.76\pm0.47$&$16.70\pm0.37$
            & $\pmb{23.72\pm0.15}$ & $21.03\pm0.43$&$21.86\pm0.68$
            & $\pmb{29.89\pm0.40}$ & $29.39\pm0.76$&$28.68\pm0.47$ \\ 
\midrule
            
500000      & $9.22\pm0.68$ & $9.12\pm0.37$&$\pmb{10.38\pm0.25}$
            &$12.53\pm0.18$ & $\pmb{12.75\pm0.24}$&$12.57\pm0.39$
            &$\pmb{17.01\pm0.34}$ & $15.95\pm0.61$&$16.46\pm0.39$ \\ 
\bottomrule
\end{tabular}}

\end{table*}

\subsection{Results on synthetic graphs}
We report the active rate under different budget $k$ of \model and other baselines on synthetic networks scaling from 10000 to 500000 generated by the powerlaw-cluster model. Since the network generation process is stochastic, we generate 100 networks for each scale and report the mean and standard deviation.
As shown in Table \ref{tab:1}, \model achieves superior performance over both IMM and GCOMB in terms of active rate. One important observation is that even we train \model on small graphs with a scale range of $30\sim50$, it still performs well on networks with several orders of magnitude larger scales. This is because \model uses the inductive graph embedding method whose parameters are independent of the network size, and for reinforcement learning, it adopts the uniform termination condition for networks with different sizes. Another observation is that the \textit{batch nodes selection} strategy can bring higher solution quality. For example, as shown in Table \ref{tab:4}, when the network scale is 10000 and 20000, \model-10 even obtains a higher active rate than \model-1. And when the network size is 500000, \model-All performs better than \model-1. This phenomenon indicates that our \textit{batch nodes selection} strategy could not only decrease the time complexity, but also enhance the effectiveness of \model. And from Fig. \ref{fig:scale} (a) we can also see that for budget $k = 30$, the running time increases linearly with regard to the network size.

\subsection{Results on real-world networks}
\begin{figure}
    \centering
    \includegraphics[width =1.0 \linewidth]{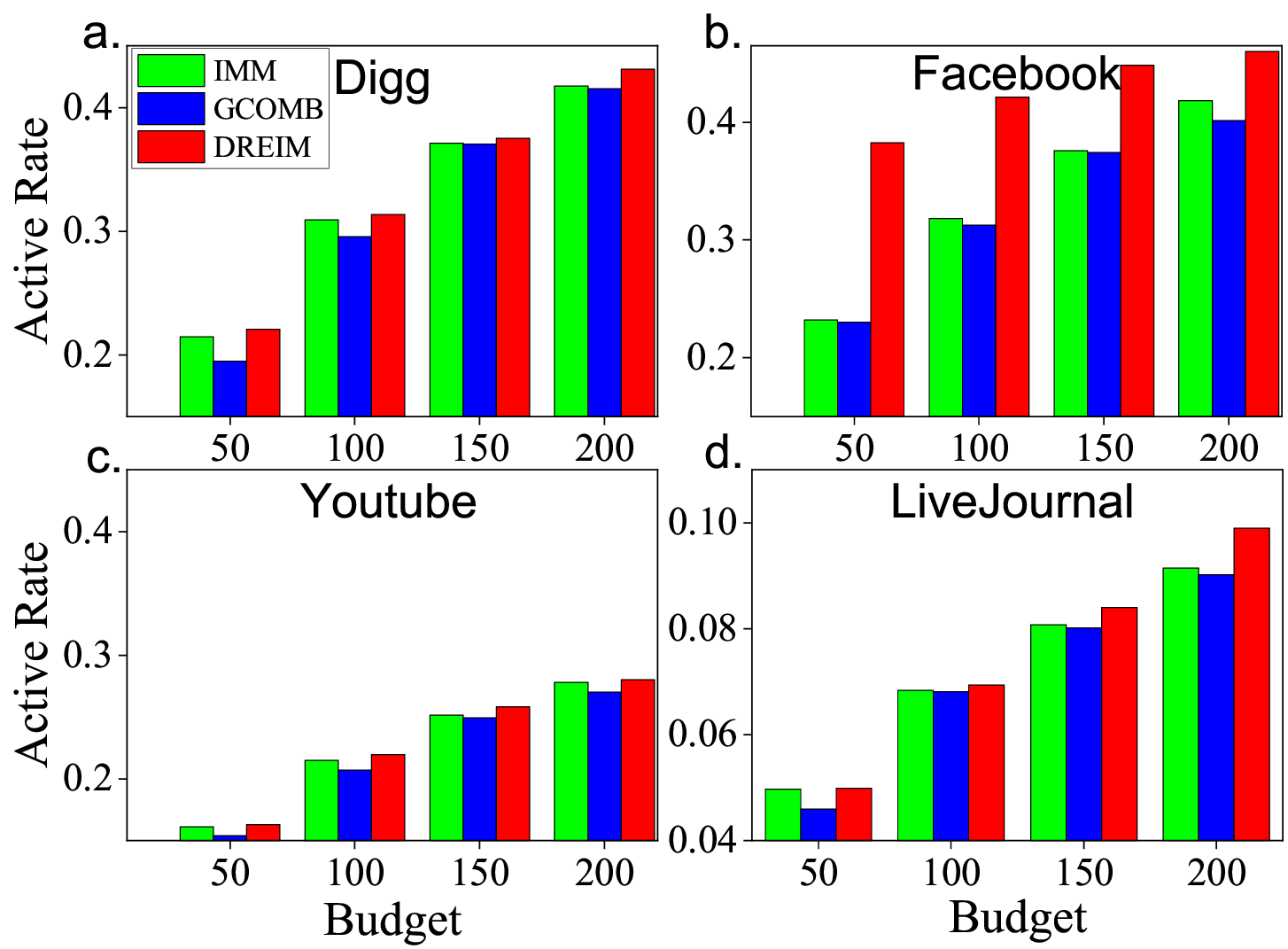}
    \caption{Active rate on real-world networks under different budget $k$.}
    \label{fig:Active Rate}
\end{figure}

\begin{figure}[ht]
    \centering
    \includegraphics[width =1.0 \linewidth]{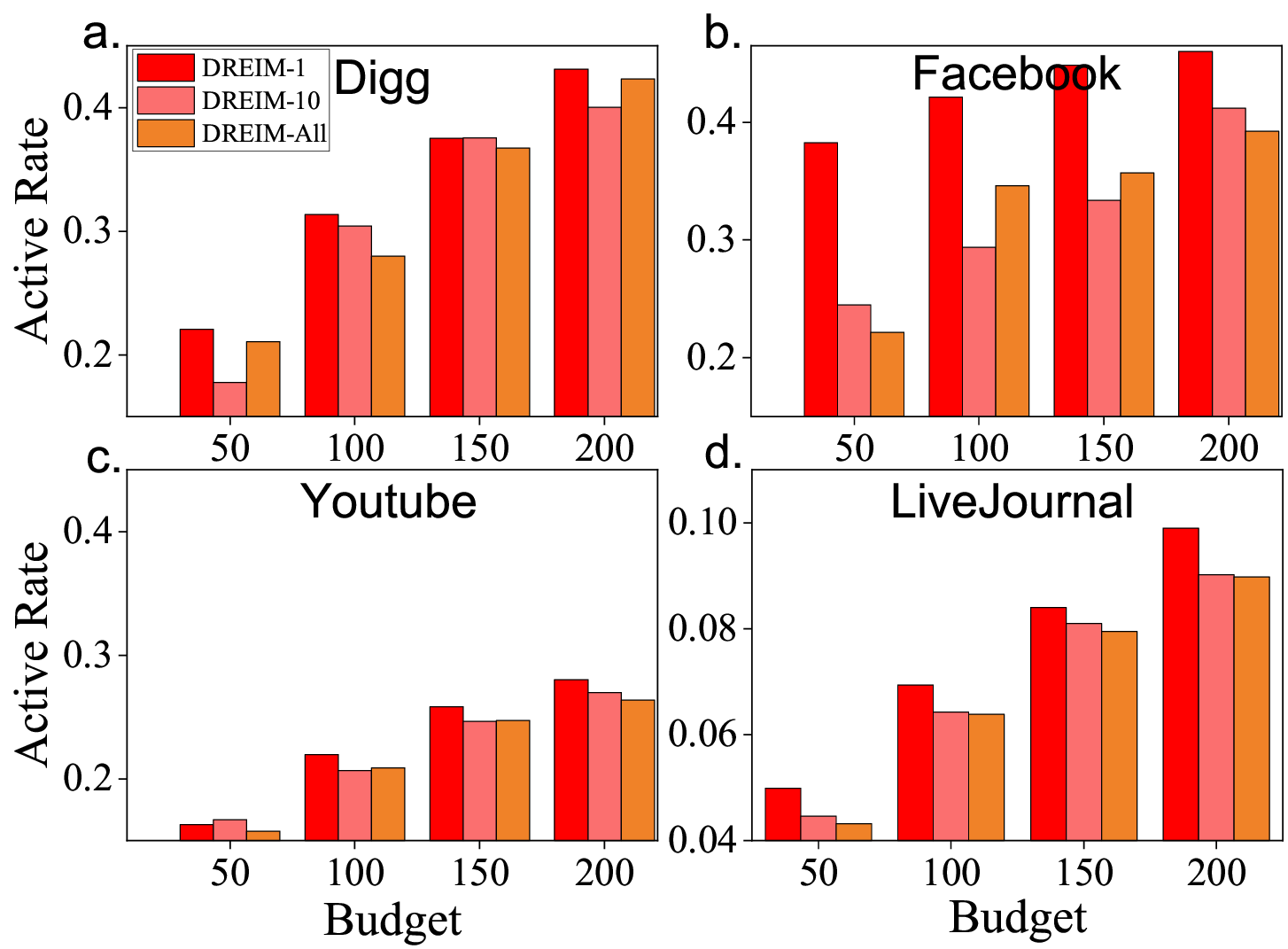}
    \caption{Active rate on real-world networks under different budget $k$ using distinct batch sizes.}
    \label{fig:batchs}
\end{figure}

In the last section, we see that \model performs well on large synthetic graphs generated by the same model as it is trained on. In this section, we test whether \model can still perform well on large real-word networks. In Fig.~\ref{fig:Active Rate} we plot the active rate curve using different methods under different budget k for every network we test. We can see that \model can always surpass all the baselines in terms of active rate for all the experiment settings, i.e., different networks and budgets. For example, when initially activating 50 nodes for Youtube, \model finally activates 16.3\% of the total nodes while IMM and GCOMB activate 16.1\% and 15.4\% of the total nodes respectively. In Fig. \ref{fig:scale} (b), we plot the wall-clock running time curve for Facebook network with $k=50$ using different batch sizes. We can see that \model-10 takes 10 times less time than \model-1 meanwhile, as shown in Fig. \ref{fig:batchs}, activates 24.5\% of the total nodes, clearly higher than 23.2\% and 23.0\% obtained by IMM and GCOMB, meaning that the running time can be dramatically reduced by exploiting the \textit{batch nodes selection strategy} and the effectiveness of \model is not greatly affected. 

\subsection{Comparison of time complexity}
As analyzed before, the time complexity of \model is $O(M+ N+N\mathrm{log}N)$, for sparse real-world networks, we can say \model has a linear time complexity $O(kM)$ when budget is k. And the empirical results in Fig. \ref{fig:scale} (a) also support our theoretical analysis.
For IMM, as reported by the authors, its time complexity is $O\left((k+\ell)(n+m) \log n / \varepsilon^{2}\right)$, i.e., IMM also has a near linear scalability with regard to the network size. For GCOMB, its time complexity is $O\left(|V|+\left|V^{g, K}\right|\left(d m_{G}+m_{G}^{2}\right)+\left|V^{g}\right| b\left(d+m_{Q}\right)\right)$, see~\cite{manchanda2020gcomb} for details. \model has a comparatively lower time complexity among all the methods. Moreover, the time complexity can be mitigated through the \textit{batch nodes selection} strategy to $O(k/bM)$ supported by results in Fig. \ref{fig:scale} (b), where $b$ is the batch size.

\begin{table}
  \caption{Hyper-parameter values.}
  \label{tab:hyper-parameters}
  \scalebox{0.8}{
  \begin{tabular}{p{2.7cm}p{1.5cm}p{5.4cm}}
    \toprule[1.5pt]
    Hyper-parameter&Value&Description\\
    \midrule[1pt]
    \small learning rate & $1\times 10^{-4}$ & the learning rate used by Adam optimizer\\
    embedding dimension & 64 & dimension of node embedding vector\\
    maximum episodes & $1\times 10^{6}$  & maximum episodes for the training process\\
    layer iterations & 3 & number of message-passing iterations\\
    reconstruction loss weight & $1\times 10^{-3}$ & weight of graph reconstruction loss \\
  \bottomrule[1.5pt]
\end{tabular}}
\end{table}
\subsection{Other settings}
All experiments are performed on a 32-core sever with 64GB memory. We conduct the validation process every 300 iterations and conduct the \textit{play game process} every 10 iterations.
For the $n$-step $Q$-learning, we set $n=5$ and 64 experiences are sampled uniformly random from the experience replay buffer. We implement the model using Tensorflow and use the Adam optimizer, and the hyper-parameters are summarized in Table \ref{tab:hyper-parameters}.

To reduce the randomness of the diffusion process, we run a large number (e.g.,~10000) of diffusion process and use the average value as the final spread value. 
\section{DISCUSSION}
In this section, we try to interpret what heuristics \model has learned. We further discuss the effect of different termination conditions on \model's performance.
\subsection{Policy analysis of \model}
In this section, we try to interpret \model's success through two observations. Firstly, the agent learns to minimize the \textit{Return}. To make it clear, we draw the inactive rate curve, as shown in Fig. \ref{fig:convergence curve} (a), where the \textit{Return} can be approximately computed by the area under the curve. Note that minimizing the \textit{Return} guides the agent to become more and more intelligent in two folds. One is that it tries to select seed nodes as few as possible to activate all the nodes, and the other one is that it prefers to select nodes that bring more decrease on inactive rate at each step. These two aspects coordinate together to guide the agent to find better influencers. Secondly, in terms of the seed nodes, \model tends to find the influencers which have a balance of connectivity based centrality, e.g., degree centrality and distance based centrality, e.g. betweenness centrality.

\subsection{Different termination conditions}
In what follows, we discuss the influences of using different termination conditions of the interaction process for training and inference. 

During the training phase, the interaction process ends when all nodes are activated, which is a uniform state for networks of different scales while during the inference phase, we use the \textit{batch nodes selection} strategy, and terminates when $|\cS| = k$, where $k$ is the budget. Previous works like~\cite{manchanda2020gcomb,nipa} adopt $|\cS| = k$ as the termination condition for both training and inference. We argue this setting will restrict \model's generalization ability because the termination state can be very different for networks of different scales. One concern is that, using different termination condition seems to be unreasonable, since traditionally researchers usually apply their model to similar problem instances. However, the empirical results in Table \ref{tab:1} and Fig. \ref{fig:Active Rate} do show that though we use different termination conditions for training and inference, we still obtain better spread quality for all budgets. We will leave exploring explanations for this phenomenon as a future direction.
\section{CONCLUSION}
In this paper, we formalize the influence maximization problem as a Markov decision process, and we design a novel reinforcement learning framework \model to tackle this problem. To our best knowledge, \model is the first work that obtains significant superior spread quality over traditional state-of-the-art method IMM. We combine the graph neural network and deep reinforcement learning together. We treat the embedding of the virtual node and the embeddings of real nodes as the state and action of the reinforcement learning. By exploiting a \textit{batch nodes selection} strategy, the computational time is greatly reduced without compromising much on solution quality.

We use synthetic graphs generated by the powerlaw-cluster model to train \model and test \model on several large synthetic and real-world social networks of different sizes from thousands to millions. And the empirical results show that \model has exceeded the performance of all the baselines.

For future work, one possible direction is to seek theoretical guarantees. Another important direction is to design more powerful and interpretable graph representation learning methods to better preserve graph information.
\section*{FUNDING}

This work was supported  in part by the National Natural Science Foundation of China (Nos. 61803248, 61872093, and U20B2051), the National Key R \& D Program of China
(No. 2018YFB1305104), Shanghai Municipal Science and Technology Major Project  (Nos.  2018SHZDZX01 and 2021SHZDZX03), ZJ Lab, and Shanghai Center for Brain Science and Brain-Inspired Technology.

\section*{DATA AVAILABILITY STATEMENT}

The data underlying this article are available in SNAP, at http://snap.stanford.edu/data.
\bibliographystyle{compj}

\begin{thebibliography}{99}

\bibitem{nguyen2016stop}
Nguyen, H.~T., Thai, M.~T., and Dinh, T.~N. (2016) Stop-and-stare: Optimal
  sampling algorithms for viral marketing in billion-scale networks.
\newblock {\em Proceedings of the International Conference on Management of
  Data},  San Francisco, California,  26 June-01 July,  pp. 695--710. ACM, New
  York.

\bibitem{chen2010scalable}
Chen, W., Wang, C., and Wang, Y. (2010) Scalable influence maximization for
  prevalent viral marketing in large-scale social networks.
\newblock {\em Proceedings of the 16th International Conference on Knowledge
  Discovery and Data Mining},  Washington, D.C.,  25-28 July,  pp. 1029--1038.
  ACM, New York.

\bibitem{Kempe}
Kempe, D., Kleinberg, J., and Tardos, E. (2003) Maximizing the spread of
  influence through a social network.
\newblock {\em Proceedings of the 9th International Conference on Knowledge
  Discovery and Data Mining},  Washington, D.C.,  24-27 August,  pp. 137--146.
  ACM, New York.

\bibitem{Tang2018sigmod}
Tang, J., Tang, X., Xiao, X., and Yuan, J. (2018) Online processing algorithms
  for influence maximization.
\newblock {\em Proceedings of the International Conference on Management of
  Data},  Houston, TX, USA,  10-15 June,  pp. 991--1005. ACM, New York.

\bibitem{newchen2010icdm}
Chen, W., Yuan, Y., and Zhang, L. (2010) Scalable influence maximization in
  social networks under the linear threshold model.
\newblock {\em Proceedings of the International Conference on Data Mining},
  Sydney, Australia,  14-17 December,  pp. 88--97. {IEEE}.

\bibitem{newcheng2013cikm}
Cheng, S., Shen, H., Huang, J., Zhang, G., and Cheng, X. (2013) Staticgreedy:
  solving the scalability-accuracy dilemma in influence maximization.
\newblock {\em Proceedings of the 22nd International Conference on Information
  and Knowledge Management},  San Francisco, California,  27 October-1
  November,  pp. 509--518. ACM, New York.

\bibitem{newcheng2014SIGIR}
Cheng, S., Shen, H., Huang, J., Chen, W., and Cheng, X. (2014) Imrank:
  Influence maximization via finding self-consistent ranking.
\newblock {\em Proceedings of the 37th International Conference on Research
  $\&$ Development in Information Retrieval},  Gold Coast, Queensland,  06-11
  July,  pp. 475--484. ACM, New York.

\bibitem{newcohen2014cikm}
Cohen, E., Delling, D., Pajor, T., and Werneck, R.~F. (2014) Sketch-based
  influence maximization and computation: Scaling up with guarantees.
\newblock {\em Proceedings of the 23rd International Conference on Information
  and Knowledge Management},  Shanghai, China,  3-7 November,  pp. 629--638.
  ACM, New York.

\bibitem{newgalhotra2015asimWWW}
Galhotra, S., Arora, A., Virinchi, S., and Roy, S. (2015) Asim: A scalable
  algorithm for influence maximization under the independent cascade model.
\newblock {\em Proceedings of the 24th International Conference on World Wide
  Web},  Florence, Italy,  18-22 May,  pp. 35--36. ACM, New York.

\bibitem{newgalhotra2016sigmod}
Galhotra, S., Arora, A., and Roy, S. (2016) Holistic influence maximization:
  Combining scalability and efficiency with opinion-aware models.
\newblock {\em Proceedings of the International Conference on Management of
  Data},  San Francisco, California,  26 June-01 July,  pp. 743--758. ACM, New
  York.

\bibitem{newgoyal2011icdm}
Goyal, A., Lu, W., and Lakshmanan, L.~V. (2011) Simpath: An efficient algorithm
  for influence maximization under the linear threshold model.
\newblock {\em Proceedings of the International Conference on Data Mining},
  Vancouver, BC, Canada,  11-14 December,  pp. 211--220. IEEE.

\bibitem{newGoyal2011VLDB}
Goyal, A., Bonchi, F., and Lakshmanan, L. V.~S. (2011) A data-based approach to
  social influence maximization.
\newblock {\em Proc. VLDB Endow.}, {\bf  5}, 73--84.

\bibitem{newJung2021icdm}
Jung, K., Heo, W., and Chen, W. (2012) Irie: Scalable and robust influence
  maximization in social networks.
\newblock {\em Proceedings of the International Conference on Data Mining},
  Brussels, Belgium,  10-13 December,  pp. 918--923. IEEE.

\bibitem{newLee2014www}
Lee, J.-R. and Chung, C.-W. (2014) A fast approximation for influence
  maximization in large social networks.
\newblock {\em Proceedings of the 23rd International Conference on World Wide
  Web},  Seoul, Korea,  7-11 April,  pp. 1157--1162. ACM, New York.

\bibitem{newNaoto2014AAAI}
Ohsaka, N., Akiba, T., Yoshida, Y., and Kawarabayashi, K.-I. (2014) Fast and
  accurate influence maximization on large networks with pruned monte-carlo
  simulations.
\newblock {\em Proceedings of the 28th AAAI Conference on Artificial
  Intelligence},  Qu\'{e}bec City, Qu\'{e}bec, Canada,  27-31 July,  pp.
  138--144. AAAI Press.

\bibitem{newSong2015parallel}
Song, G., Zhou, X., Wang, Y., and Xie, K. (2015) Influence maximization on
  large-scale mobile social network: A divide-and-conquer method.
\newblock {\em IEEE Trans. Parallel Distrib. Syst.}, {\bf  26}, 1379--1392.

\bibitem{newTang2017ASONAM}
Tang, J., Tang, X., and Yuan, J. (2017) Influence maximization meets efficiency
  and effectiveness: A hop-based approach.
\newblock {\em Proceedings of the International Conference on Advances in
  Social Networks Analysis and Mining},  Sydney, Australia,  31 July-03 August,
   pp. 64--71. ACM, New York.

\bibitem{newzhou2013icdm}
Zhou, C., Zhang, P., Guo, J., Zhu, X., and Guo, L. (2013) Ublf: An upper bound
  based approach to discover influential nodes in social networks.
\newblock {\em Proceedings of the International Conference on Data Mining},
  Dallas, TX, USA,  7-10 December,  pp. 907--916. IEEE.

\bibitem{newzhou2014www}
Zhou, C., Zhang, P., Guo, J., and Guo, L. (2014) An upper bound based greedy
  algorithm for mining top-k influential nodes in social networks.
\newblock {\em Proceedings of the 23rd International Conference on World Wide
  Web},  Seoul, Korea,  7-11 April,  pp. 421--422. ACM, New York.

\bibitem{leskovec2007costcelf}
Leskovec, J., Krause, A., Guestrin, C., Faloutsos, C., VanBriesen, J., and
  Glance, N. (2007) Cost-effective outbreak detection in networks.
\newblock {\em Proceedings of the 13th International Conference on Knowledge
  Discovery and Data Mining},  San Jose, California,  12-15 August,  pp.
  420--429. ACM, New York.

\bibitem{2012Maximizing}
Borgs, C., Brautbar, M., Chayes, J.~T., and Lucier, B. (2014) Maximizing social
  influence in nearly optimal time.
\newblock In Chekuri, C. (ed.), {\em Proceedings of the 25th Annual {ACM-SIAM}
  Symposium on Discrete Algorithms},  Portland, Oregon,  5-7 January,  pp.
  946--957. {SIAM}.

\bibitem{tang2014influence}
Tang, Y., Xiao, X., and Shi, Y. (2014) Influence maximization: Near-optimal
  time complexity meets practical efficiency.
\newblock {\em Proceedings of the International Conference on Management of
  Data},  Snowbird, Utah, USA,  22-27 June,  pp. 75--86. ACM, New York.

\bibitem{tang2015influenceIMM}
Tang, Y., Shi, Y., and Xiao, X. (2015) Influence maximization in near-linear
  time: A martingale approach.
\newblock {\em Proceedings of the International Conference on Management of
  Data},  Melbourne, Victoria, Australia,  31 May-04 June,  pp. 1539--1554.
  ACM, New York.

\bibitem{bucur2016evelutionIM}
Bucur, D. and Iacca, G. (2016) Influence maximization in social networks with
  genetic algorithms.
\newblock In Squillero, G. and Burelli, P. (eds.), {\em Applications of
  Evolutionary Computation},  Cham,  30 March-1 April,  pp. 379--392. Springer
  International Publishing.

\bibitem{centralityheuristic1}
Tabak, B.~M., Takami, M., Rocha, J.~M., Cajueiro, D.~O., and Souza, S.~R.
  (2014) Directed clustering coefficient as a measure of systemic risk in
  complex banking networks.
\newblock {\em Physica A}, {\bf  394}, 211--216.

\bibitem{centralityheuristic2}
Wilson, C., Boe, B., Sala, A., Puttaswamy, K.~P., and Zhao, B.~Y. (2009) User
  interactions in social networks and their implications.
\newblock {\em Proceedings of the 4th European Conference on Computer Systems},
   Nuremberg, Germany,  1-3 April,  pp. 205--218. ACM, New York.

\bibitem{2009Efficient}
Chen, W., Wang, Y., and Yang, S. (2009) Efficient influence maximization in
  social networks.
\newblock {\em Proceedings of the 15th International Conference on Knowledge
  Discovery and Data Mining},  Paris, France,  28 June-1 July,  pp. 199--208.
  ACM, New York.

\bibitem{dai2017learning}
Dai, H., Khalil, E.~B., Zhang, Y., Dilkina, B., and Song, L. (2017) Learning
  combinatorial optimization algorithms over graphs.
\newblock {\em Proceedings of the 31st International Conference on Neural
  Information Processing Systems},  Long Beach, California,  4-9 December,  pp.
  6351--6361. Curran Associates Inc., Red Hook, NY, USA.

\bibitem{manchanda2020gcomb}
Manchanda, S., MITTAL, A., Dhawan, A., Medya, S., Ranu, S., and Singh, A.
  (2020) Gcomb: Learning budget-constrained combinatorial algorithms over
  billion-sized graphs.
\newblock {\em Proceedings of the 34th International Conference on Neural
  Information Processing Systems},  Online Event,  6-12 December,  pp.
  20000--20011. Curran Associates, Inc., Red Hook, NY, USA.

\bibitem{li2018guidedtreesearch}
Li, Z., Chen, Q., and Koltun, V. (2018) Combinatorial optimization with graph
  convolutional networks and guided tree search.
\newblock {\em Proceedings of the 32nd International Conference on Neural
  Information Processing Systems},  Montr\'{e}al, Canada,  3-8 December,  pp.
  537--546. Curran Associates Inc., Red Hook, NY, USA.

\bibitem{richardson2002mining_imfirst1}
Richardson, M. and Domingos, P. (2002) Mining knowledge-sharing sites for viral
  marketing.
\newblock {\em Proceedings of the 8th International Conference on Knowledge
  Discovery and Data Mining},  Edmonton, Alberta, Canada,  23-26 July,  pp.
  61--70. ACM, New York.

\bibitem{domingos2001mining_imfirst2}
Domingos, P. and Richardson, M. (2001) Mining the network value of customers.
\newblock {\em Proceedings of the International 7th Conference on Knowledge
  Discovery and Data Mining},  San Francisco, California,  26-29 August,  pp.
  57--66. ACM, New York.

\bibitem{arora2017debunking}
Arora, A., Galhotra, S., and Ranu, S. (2017) Debunking the myths of influence
  maximization: An in-depth benchmarking study.
\newblock {\em Proceedings of the International Conference on Management of
  Data},  Chicago, Illinois, USA,  14-19 May,  pp. 651--666. ACM, New York.

\bibitem{sen2008collectivenodeclassification}
Sen, P., Namata, G., Bilgic, M., Getoor, L., Galligher, B., and Eliassi-Rad, T.
  (2008) Collective classification in network data.
\newblock {\em AI Mag.}, {\bf  29}, 93.

\bibitem{liben2007linkprediction}
Liben-Nowell, D. and Kleinberg, J. (2003) The link prediction problem for
  social networks.
\newblock {\em Proceedings of the 12th International Conference on Information
  and Knowledge Management},  New York, NY,  2-8 November,  pp. 556--559. ACM,
  New York.

\bibitem{herman2000graphvisualization}
Herman, I., Melan\c{c}on, G., and Marshall, M.~S. (2000) Graph visualization
  and navigation in information visualization: A survey.
\newblock {\em IEEE Trans. Vis. Comput. Graph.}, {\bf  6}, 24--43.

\bibitem{fan2019drbc}
Fan, C., Zeng, L., Ding, Y., Chen, M., Sun, Y., and Liu, Z. (2019) Learning to
  identify high betweenness centrality nodes from scratch: A novel graph neural
  network approach.
\newblock {\em Proceedings of the 28th International Conference on Information
  and Knowledge Management},  Beijing, China,  3-7 November,  pp. 559--568.
  ACM, New York.

\bibitem{kipf2017semisupervised}
Kipf, T.~N. and Welling, M. (2017) Semi-supervised classification with graph
  convolutional networks.
\newblock {\em Proceedings of the 5th International Conference on Learning
  Representations},  Toulon, France,  24-26 April. {OpenReview}.net.

\bibitem{hamilton2017inductive}
Hamilton, W.~L., Ying, R., and Leskovec, J. (2017) Inductive representation
  learning on large graphs.
\newblock {\em Proceedings of the 31st International Conference on Neural
  Information Processing Systems},  Long Beach, California,  4-9 December,  pp.
  1025--1035. Curran Associates Inc., Red Hook, NY, USA.

\bibitem{sutton1988reinforcement}
Sutton, R.~S. and Barto, A.~G. (2018) {\em Reinforcement learning: An
  introduction}. {MIT} press.

\bibitem{bengio2020machine4optimization}
Bengio, Y., Lodi, A., and Prouvost, A. (2021) Machine learning for
  combinatorial optimization: A methodological tour d'horizon.
\newblock {\em Eur. J. Oper. Res.}, {\bf  290}, 405--421.

\bibitem{bello2016neurale2e1}
Bello, I., Pham, H., Le, Q.~V., Norouzi, M., and Bengio, S. (2016) Neural
  combinatorial optimization with reinforcement learning.
\newblock {\em CoRR}, {\bf  abs/1611.09940}.

\bibitem{selsam2018learninge2e2}
Selsam, D., Lamm, M., B\"{u}nz, B., Liang, P., de Moura, L., and Dill, D.~L.
  (2019) Learning a {SAT} solver from single-bit supervision.
\newblock {\em Proceedings of the 7th International Conference on Learning
  Representations},  New Orleans, LA, USA,  6-9 May. OpenReview.net.

\bibitem{bonami2018learningconfigure}
Bonami, P., Lodi, A., and Zarpellon, G. (2018) Learning a classification of
  mixed-integer quadratic programming problems.
\newblock In van Hoeve, W.-J. (ed.), {\em Integration of Constraint
  Programming, Artificial Intelligence, and Operations Research},  Delft, The
  Netherlands,  26-29 June,  pp. 595--604. Springer International Publishing.

\bibitem{wilder2019meldingconfigure2}
Wilder, B., Dilkina, B., and Tambe, M. (2019) Melding the data-decisions
  pipeline: Decision-focused learning for combinatorial optimization.
\newblock {\em Proceedings of the 23rd AAAI Conference on Artificial
  Intelligence},  Honolulu, Hawaii, USA,  27 January-1 February,  pp.
  1658--1665. {AAAI} Press.

\bibitem{lodi2017learningalongside1}
Lodi, A. and Zarpellon, G. (2017) On learning and branching: a survey.
\newblock {\em TOP}, {\bf  25}, 207--236.

\bibitem{cappart2019improvingalongside2}
Cappart, Q., Goutierre, E., Bergman, D., and Rousseau, L.-M. (2019) Improving
  optimization bounds using machine learning: Decision diagrams meet deep
  reinforcement learning.
\newblock {\em Proceedings of the 23rd AAAI Conference on Artificial
  Intelligence},  Honolulu, Hawaii, USA,  27 January-1 February,  pp.
  1443--1451. {AAAI} Press.

\bibitem{fan2020finder}
Fan, C., Zeng, L., Sun, Y., and Liu, Y.-Y. (2020) Finding key players in
  complex networks through deep reinforcement learning.
\newblock {\em Nat. Mach. Intell.}, {\bf  2}, 317--324.

\bibitem{BA}
Barabási, A.-L. and Albert, R. (1999) Emergence of scaling in random networks.
\newblock {\em Science}, {\bf  286}, 509--512.

\bibitem{1998WS}
Watts, D.~J. and Strogatz, S.~H. (1998) Collective dynamics of `small-world'
  networks.
\newblock {\em Nature}, {\bf  393}, 440--442.

\bibitem{ER}
Erd\"{o}s, P. and R\'{e}nyi, A. (1959) On random graphs i.
\newblock {\em Publ. Math.-Debr.}, {\bf  6}, 290.

\bibitem{powerlaw-cluster}
Holme, P. and Kim, B.~J. (2002) Growing scale-free networks with tunable
  clustering.
\newblock {\em Phys. Rev. E}, {\bf  65}, 026107.

\bibitem{nipa}
Sun, Y., Wang, S., Tang, X., Hsieh, T.-Y., and Honavar, V. (2020) Adversarial
  attacks on graph neural networks via node injections: A hierarchical
  reinforcement learning approach.
\newblock {\em Proceedings of The Web Conference},  Taipei, Taiwan,  20-24
  April,  pp. 673--683. ACM, New York.

\bibitem{fittedQ}
Riedmiller, M. (2005) Neural fitted q iteration -- first experiences with a
  data efficient neural reinforcement learning method.
\newblock In Gama, J., Camacho, R., Brazdil, P.~B., Jorge, A.~M., and Torgo, L.
  (eds.), {\em Machine Learning: ECML},  Porto, Portugal,  3-7 October,  pp.
  317--328. Springer Berlin Heidelberg.

\bibitem{snapnets}
Leskovec, J. and Krevl, A. (2014).
\newblock {SNAP Datasets}: {Stanford} large network dataset collection.
\newblock \url{http://snap.stanford.edu/data}.

\end{thebibliography}

\end{document}